\documentclass[12pt]{article}
\usepackage{epsfig}

\makeatletter
\renewcommand\section{\@startsection {section}{1}{\z@}%
                                   {-3.5ex \@plus -1ex \@minus -.2ex}
                                   {2.3ex \@plus.2ex}%
                                   {\normalfont\large\bfseries}}
\renewcommand\subsection{\@startsection{subsection}{2}{\z@}%
                                     {-3.25ex\@plus -1ex \@minus -.2ex}%
                                     {1.5ex \@plus .2ex}%
                                     {\normalfont\bfseries}}
\makeatother

\def\IZ{\relax\ifmmode\mathchoice
{\hbox{\cmss Z\kern-.4em Z}}{\hbox{\cmss Z\kern-.4em Z}}
{\lower.9pt\hbox{\cmsss Z\kern-.4em Z}} {\lower1.2pt\hbox{\cmsss
Z\kern-.4em Z}}\else{\cmss Z\kern-.4em Z}\fi}
\def\IR{\relax{\rm I\kern-.18em R}}

\def\one{{\hbox{ 1\kern-.8mm l}}}

\def\tr{{\rm tr\,}}

\newlength{\bredde}
\def\slash#1{\settowidth{\bredde}{$#1$}\ifmmode\,\raisebox{.15ex}{/}
\hspace*{-\bredde} #1\else$\,\raisebox{.15ex}{/}\hspace*{-\bredde}
#1$\fi}

\newsavebox{\zzzbar}
\sbox{\zzzbar}
  {\setlength{\unitlength}{0.9em}
  \begin{picture}(0.6,0.7)
  \thinlines
  \put(0,0){\line(1,0){0.6}}
  \put(0,0.75){\line(1,0){0.575}}
  \multiput(0,0)(0.0125,0.025){30}{\rule{0.3pt}{0.3pt}}
  \multiput(0.2,0)(0.0125,0.025){30}{\rule{0.3pt}{0.3pt}}
  \put(0,0.75){\line(0,-1){0.15}}
  \put(0.015,0.75){\line(0,-1){0.1}}
  \put(0.03,0.75){\line(0,-1){0.075}}
  \put(0.045,0.75){\line(0,-1){0.05}}
  \put(0.05,0.75){\line(0,-1){0.025}}
  \put(0.6,0){\line(0,1){0.15}}
  \put(0.585,0){\line(0,1){0.1}}
  \put(0.57,0){\line(0,1){0.075}}
  \put(0.555,0){\line(0,1){0.05}}
  \put(0.55,0){\line(0,1){0.025}}
  \end{picture}}
\newcommand{\Tr}{\mathop{\mbox{Tr}}\nolimits}

\newcommand {\n}{\nonumber \\}

\newcommand{\ena}{\end{eqnarray}}
\newcommand{\bea}{\begin{eqnarray}}
\newcommand{\eea}{\end{eqnarray}}

\usepackage{graphicx}

\newcommand{\beq}{\begin{equation}}
\newcommand{\eeq}{\end{equation}}
\newcommand{\ber}{\begin{array}}
\newcommand{\eer}{\end{array}}

\newcommand{\del}{\partial}

\newcommand{\dsty}{\displaystyle}

\newcommand{\de}{\delta}

\newcommand{\eps}{\varepsilon}

\usepackage{amsmath}
\usepackage{amssymb}

\begin{document}
\begin{titlepage}
\begin{flushright}
\phantom{arXiv:yymm.nnnn}
\end{flushright}
\vfill
\begin{center}
{\Large\bf Randomized Wilson loops, reduced models\vspace{2mm}\\
and the large $D$ expansion}    \\
\vskip 15mm
Oleg Evnin
\vskip 10mm
{\em Institute of Theoretical Physics, Academia Sinica\\
Zh\=onggu\=anc\=un d\=ongl\`u 55, Beijing 100190, China}
\vskip 3mm
{\small\noindent  {\tt eoe@itp.ac.cn}}
\end{center}
\vfill

\begin{center}
{\bf ABSTRACT}\vspace{3mm}
\end{center}

Reduced models are matrix integrals believed to be related to the large $N$ limit of gauge theories. These integrals are known to simplify further when the number of matrices $D$ (corresponding to the number of space-time dimensions in the gauge theory) becomes large.
Even though this limit appears to be of little use for computing the standard
rectangular Wilson loop (which always singles out two directions out of $D$), a meaningful
large $D$ limit can be defined for a randomized Wilson loop (in which all $D$ directions
contribute equally). In this article, a proof-of-concept demonstration of this approach
is given for the simplest reduced model (the original Eguchi-Kawai model) and the simplest
randomization of the Wilson loop (Brownian sum over random walks). The resulting averaged Wilson
loop displays a scale behavior strongly reminiscent of the area law.

\vfill

\end{titlepage}

\section{Introduction}

Reduced models (for a textbook exposition, see \cite{makeenko}, for a review of more recent developments, see \cite{Azeyanagi:2010ne}) are matrix integrals
whose large $N$ limit is related to large $N$ limits of gauge theories. The simplest and historically the first of these models, named after Eguchi and Kawai \cite{ek}, is simply obtained
by a full dimensional reduction of a pure gauge theory to zero dimensions. The corresponding action
reads
\beq
S = - \frac{1}{4 g^2} \tr ( [A_\mu, A_\nu ]^2 )
\equiv - \frac{1}{4 g^2} \lambda ^{abcd}
A_\mu^{a}A_\mu^{b}A_\nu^{c}A_\nu^{d},
\label{originalaction}
\eeq
where
\beq
\lambda ^{abcd} =  \frac{1}{4} \left\{ \tr ( [t^a,t^c][t^b,t^d])
+ (a\leftrightarrow b)
+ (c\leftrightarrow d)
+ \left( \begin{array}{c}
a\leftrightarrow b  \\
c\leftrightarrow d
\end{array}
\right)  \right\}.
\eeq
Here, the Greek indices label the $D$ dimensions (of the original gauge theory space-time) and the Roman indices label the adjoint representation of the gauge group $SU(N)$ with generators $t^a$. Various quantities are obtained by taking ``path integrals'' (which are ordinary integrals
for this zero-dimensional case):
\beq
\langle f(A)\rangle=\int dA_\mu^a\,f(A) e^{-S[A]}.
\eeq
Of particular interest is the reduced model analog of the Wilson loop:
\bea
W[\xi(s)]&=&\int dA_\mu^a \,\frac1N\tr\mbox{\rm\bf P}\exp\left[i\int\limits_0^{L_0} ds\, \dot\xi^\mu(s)A_\mu^a t_a\right] e^{-S[A]}\vspace{3mm}\n
&\equiv&
\int dA_\mu^a\,e^{-S[A]}\sum\limits_{n=0}^\infty (i)^n \int\limits_0^{L_0} ds_1\int\limits_{s_1}^{L_0}ds_2\cdots \int\limits_{s_{n-1}}^{L_0}ds_n \frac1N\tr\prod\limits_{k=1}^n \dot\xi^{\mu_k}(s_k)A_{\mu_k}^{a_k} t_{a_k}.
\label{wilson}
\eea

Even though the Eguchi-Kawai model was originally hoped to reproduce large $N$
properties of the gauge theory from which it is obtained by dimensional reduction,
it later turned out that this equivalence is undermined by a spontaneous breakdown
of the $R^D$ symmetry that shifts the values of the matrices $A$ by matrices proportional
to unity. Variations of the Eguchi-Kawai model (the so-called twisted \cite{twisted}
and quenched \cite{quenched} models) have been proposed to reproduce the large $N$ limit of the gauge theory (in particular, the reduced model Wilson loop of the type (\ref{wilson}) would actually equal the gauge theory Wilson loop in the large $N$ limit).
However, they have been subsequently shown to suffer from non-perturbative problems \cite{problems}. A new generation of reduced models has been subsequently developed \cite{Azeyanagi:2010ne,current}, and it is believed to reproduce correctly the properties infinite volume gauge theories. All these more sophisticated reduced models are closely related to the
original Eguchi-Kawai model, even if computationally less tractable.

In this paper, all derivations will be performed in the context of the Eguchi-Kawai
model. The ideas are nevertheless quite general and should apply to more realistic
reduced models as well. This, however, would require a systematic development of the large $D$
techniques (see below) for a broader class of reduced models. Likewise, introducing
supersymmetry may protect the above-mentioned $R^D$ symmetry from spontaneous breaking,
and would make the straightforward reduced model obtained by dimensional reduction
of the gauge theory describe the large $N$ limit (see \cite{super}, for recent considerations
in the context of string theory, see \cite{string}).

A further significant simplification occurs in reduced models when the number of matrices
$D$ (corresponding to the number of space-time dimensions of the original gauge theory)
becomes large. For the Eguchi-Kawai model, this simplification can be demonstrated by the following technique \cite{hnt}. (The large $D$ techniques have been recently applied to matrix quantum mechanics \cite{morita} as well, though matrix integrals will remain our main focus in the present paper.) One first introduces an auxiliary field $h_{ab}$,
which is a real symmetric tensor, with the following action.
\bea
S[A,h] & =& \frac{1}{4 g^2} \lambda ^{abcd}
( h_{ab} h_{cd} - A_\mu^{a}A_\mu^{b} h_{cd} 
-  h_{ab} A_\nu^{c}A_\nu^{d} ) \\
&=& \frac{1}{4 g^2} \lambda ^{abcd}
h_{ab} h_{cd} + \frac{1}{2g} K_{ab} A_\mu^{a}A_\mu^{b},
\label{auxiliaryaction}
\eea
where $K$ is a dimensionless kernel
\beq
 K_{ab} = - \frac{1}{g} \lambda ^{abcd} h_{cd}
\eeq
for the Gaussian quadratic form of $A_\mu^{a}$.
Integrating out the auxiliary field, one reproduces the original
action (\ref{originalaction}).

Since (\ref{auxiliaryaction}) is quadratic in $A_\mu ^a$, it can be integrated out first.
The propagator for $A_\mu^a$ is given by
\beq
\langle A_\mu^a A_\nu^{b} \rangle
= g \delta_{\mu\nu} (K^{-1})_{ab} .
\label{Apropgeneral}
\eeq
By rescaling the $h$ as
$\tilde{h}_{ab} = h_{ab}/g \sqrt{D}$,
which is now dimensionless, we can rewrite the effective action
as
\beq
S_{\mbox{\scriptsize eff}}
=
 \frac{D}{2} \left \{ \Tr \ln K + 
\frac{1}{2} \lambda ^{abcd} \tilde{h}_{ab} \tilde{h}_{cd} \right \} .
\eeq
Therefore, in the large $D$ limit, the integration over $\tilde{h}$ 
is dominated by the saddle point and one can perform
a systematic $1/D$ expansion.
The saddle point equation can be given as
\beq
(K^{-1})_{cd} \frac{\delta K_{dc}}{\delta \tilde{h}_{ab}}
+ \lambda ^{abcd} \tilde{h}_{cd} = 0 .
\label{speq}
\eeq
We denote the relevant saddle point $\tilde{h}_{ab}^{(0)}$,
and if one assumes that it preserves the SU($N$) symmetry,
$\tilde{h}_{ab}^{(0)} = v \delta_{ab}$.
From the saddle point equation (\ref{speq}),
we obtain $v=1/\sqrt{2N}$, i.e.,
\beq
\tilde{h}_{ab} = \frac{1}{\sqrt{2N}} \delta_{ab}.
\eeq
The saddle point values of $K$ and its inverse (which is all one needs for computations to
leading order in $1/D$) are
\beq
K_{ab}= \sqrt{2ND} \delta_{ab},\qquad (K^{-1})_{ab} = \frac{1}{\sqrt{2ND}}\delta_{ab}.
\eeq
Then, from (\ref{Apropgeneral}), to compute the value of a certain observable to
leading order in $1/D$, one simply needs to contract all the $A$'s with
\beq
\langle A_\mu^a A_\nu^{b} \rangle
= \frac{g}{\sqrt{2ND}}\,\delta_{\mu\nu} \delta_{ab}.
\label{Acontract}
\eeq
and sum over all such contractions.

It is interesting to see what these techniques have to say about the expectation
value of the Wilson loop (\ref{wilson}), a crucial observable for gauge theories.
It may come as a disappointment that there is no non-trivial large $D$ limit for
any fixed Wilson loop (i.e., Wilson loop for a given contour). A simple fixed
Wilson loop singles out directions in the space of the $A_\mu^a$'s. On the contrary,
in a theory with a large number of degrees of freedom (such as a large $D$ limit
of reduced models), simplifications occur for observables averaged over a significant fraction
of these degrees of freedom, not for observables to which just a few degrees of freedom contribute.

Mathematically, one can try to take a na\"\i ve large $D$ limit of the Wilson loop (\ref{wilson}). For simplicity, one can take the contour to be rectangular in the $(1,2)$-plane.
Then, only powers of $A_1$ and $A_2$ appear in the expansion of the path-ordered exponential
in (\ref{wilson}). According to (\ref{Acontract}) contractions of $A_1$ and $A_2$ will
always produce factors of $g/\sqrt{2ND}$. On the other hand, organizing the large $N$ and $D$ limit
as a power expansion in $1/N$ and $1/D$ requires keeping the modified 't Hooft coupling $g^2ND$ fixed \cite{hnt}. Hence, any combination of powers of $A_1$ and $A_2$ will be suppressed by
powers of $1/D$ in the large $D$ limit, and the $D\to\infty$ limit of the simple rectangular
Wilson loop will equal 1, containing no information on the dynamics of the theory.
It is only when one considers combinations of the sort $A_\mu A_\mu$ that contracting the
$\mu$ and $\nu$ indices in (\ref{Acontract}) lends an additional factor of $D$, and hence
a non-trivial $D\to\infty$ limit. This is in agreement with the heuristic observations
of the previous paragraph.

There is a natural modification of the traditional Wilson loop observable that does not
suffer from the lack of a meaningful and tractable large $D$ limit. Namely, one can perform
averaging of the Wilson loop over a large number of different contours. If the chosen 
ensemble of contours is rich enough, many directions (many different $A_\mu^a$'s) will contribute, and the large $D$ limit will become dynamically non-trivial. In this paper,
we shall consider the simplest ensemble of uncorrelated random walks (Brownian motion).
However, the idea is quite general and should apply to more complex ensembles of loops as well.

The Brownian randomized Wilson loops are certainly quite different from the regular Wilson loops familiar to the reader. For one thing, the typical fractal dimension \cite{ambjorn} of a random walk trajectory is 2 (and it is nowhere differentiable). Still, one can give two reasons why considering sums over such ensembles of contours is a reasonable idea. First, (in a Euclidean theory) real trajectories of quantum particles (quarks) have fractal dimension 2 (and not 1 like the
idealized infinitely heavy quark implicit in the usual Wilson loop). Second, the difference
between regular and random Wilson loops appears to diminish in a sense for a large number of dimensions (which is the case that concerns us here). To clarify this, one can consider a
random walk with $L$ steps on a $D$-dimensional hypercubic lattice. This walk would occupy
a volume with $L^{D/2}$ points, whereas the walk itself contains $L$ points, so (since
there are no long range correlations) one can estimate the fraction of self-intersections
(relatively to the total number of points) as $L/L^{D/2}$. So, of course, in two dimensions
self-intersections typically occupy a finite fraction of the entire trajectory and there
is nothing like a notion of the perimeter and interior of the loop. Yet, in a very large number
of dimensions, intersections become progressively more rare, and one should expect
the random Wilson loops to look more and more tame.

In what follows, we shall consider two cases of Wilson loop randomization. First, we shall generate the Wilson loop contour by a completely unconstrained random walk. In this case,
the contour does not necessarily start and end at the same point, so a gauge theory
interpretation of such a `loop' is questionable (as it is not gauge invariant).
However, the computation is particularly simple for this case, hence it is
a nice toy model to develop the relevant techniques. Thereafter, we shall consider
a sum over closed Wilson loops generated by a random walk constrained to return to its origin.

\section{Open Wilson loops (a toy example)}

We shall now proceed implementing the Wilson loop randomization program outlined in the introduction. To this end, we define the Brownian average of the Wilson loop as
\beq
\begin{array}{l}
\dsty W_{Br}(L)=\int d\xi(s) \de[\dot\xi^2-1]\int dA_\mu^a \,\frac1N\tr\mbox{\rm\bf P}\exp\left[i\int\limits_0^{L^2/\eps} ds\, \dot\xi^\mu(s)A_\mu^a t_a\right] e^{-S[A]}\vspace{3mm}\\
\dsty\hspace{2mm}\equiv
\int d\xi(s) \de[\dot\xi^2-1]\int dA_\mu^a e^{-S[A]}\sum\limits_{n=0}^\infty i^n \int\limits_0^{L^2/\eps} ds_1\int\limits_{s_1}^{L^2/\eps}ds_2\cdots \int\limits_{s_{n-1}}^{L^2/\eps}ds_n \frac1N\tr\prod\limits_{k=1}^n \dot\xi^{\mu_k}(s_k)A_{\mu_k}^{a_k} t_{a_k}.
\end{array}
\label{Brwilson}
\eeq
The measure of the integration over paths $\xi(s)$ can be straightforwardly understood:
we are integrating over all functions $\xi(s)$ subject to the constraint $\de[\dot\xi^2-1]$
enforcing $s$ to be the proper length parameter of the path. Note that we have replaced
the total length of the path $L_0$ by a formally divergent expression $L^2/\eps$, where $\eps$ is a cut-off associated with integrating over $\xi(s)$. This is essential to maintain
a meaningful continuum limit (as we shall see below explicitly). Indeed, random paths of a finite total length collapse to a point in the continuum limit. It is only paths of length
$L^2/\eps$ that occupy a finite region of space (whose size is of order $L$) as $\eps$ is taken to 0. This is immediately related to the fractal dimension of a random walk being equal
to 2.

The path integral over $\xi$ in (\ref{Brwilson}) may appear complicated due to the non-linear constraint imposed by $\de[\dot\xi^2-1]$. This complexity is, however, spurious.
The key idea for taking such path integrals over random walks is that the measure is
completely factorized ($\de[\dot\xi^2-1]$ is a product of $\de$-functions at each point)
and has the form of a factorized probability distribution in the celebrated central limit
theorem of probability theory (with the limit of a large number of independent random variables being simply the continuum limit of the path integral). Then the result of the path integration (at least for observables depending on integrals of $\dot\xi$ over finite intervals, rather than
values of $\dot\xi$ at a point) does not depend in the continuum limit on the precise form of the distribution
of $\dot\xi$, but only on the variance of this distribution. Hence, $\de[\dot\xi^2-1]$
can be replaced by a product of Gaussian distributions, which is straightforwardly treated
in a path integral.

There are many ways to implement the measure transformation we have just described.
For example, \cite{polyakov} transforms the measure $\int d\xi(s)\de[\dot\xi^2-1]$ using a Lagrange multiplier before invoking the central limit theorem. Here, we shall take a more straightforward route, manipulating the continuum limit of the measure in (\ref{Brwilson}) directly. (An alternative justification for all these formal manipulations with the path integral can be obtained, for example, by considering the continuum limit of random walks on a hypercubic lattice, which is an exactly solvable problem.)

Let us first consider (for simplicity) the expectation value of a general observable depending
on $\xi$ (integral of $\dot\xi$), but not on the values of $\dot\xi$ at a point (note that (\ref{Brwilson}) is not quite of that form, but similar ideas apply, as we shall discuss below):
\beq
\langle F\rangle= \int d\xi(s) \de[\dot\xi^2-1] F(\xi(s_1),\cdots,\xi(s_k)).
\label{F}
\eeq
(We shall assume $s_1\le\cdots\le s_k$.) Generically, one can expand $F$ in a power
series with terms of the form
\beq
(\xi(s_k))^{a_k}\cdots (\xi(s_1))^{a_1}.
\eeq
Using $\xi(s_i)=(\xi(s_i)-\xi(s_{i-1}))+(\xi(s_{i-1})-\xi(s_{i-2}))+\cdots+(\xi(s_1)-\xi(0))$ (we assume that the random walk starts at $\xi(0)=0$), it can also be re-written as a sum of terms of the form
\beq
(\xi(s_k)-\xi(s_{k-1}))^{b_k}\cdots (\xi(s_1)-\xi(s_0))^{b_1}\equiv \left(\int\limits_{s_k}^{s_{k-1}}ds\,\dot\xi\right)^{b_k}\cdots \left(\int\limits_{s_1}^{s_0}ds\,\dot\xi\right)^{b_1}.
\eeq
Since the measure in (\ref{F}) is pointwise factorized with respect to $\dot\xi$, and the factors in the last expression depend on $\dot\xi(s)$ in non-intersecting ranges, the expectation value will likewise factorize (with an appropriate normalization of the path integral). It is therefore enough to consider the expectation value of just one factor:
\beq
I(s,s')=\int dv(s) \de[v^2(s)-1]\left(\int\limits_{s}^{s'}ds\,v(s)\right)^{b},
\label{Iss}
\eeq
where we have renamed $\dot\xi$ to $v$. All the other observables of the type we are presently considering will be expressed as sums of products of such expectation values.

As was to be expected, (\ref{Iss}) is again of the form characteristic of the central limit theorem: a large (infinite in the continuum limit) number of uncorrelated variables with identical distributions, only appearing as an averaged combination in the observable to be studied. To study the continuum limit more explicitly, it is convenient to discretize the path integral in (\ref{Iss}), splitting the interval $(s,s')$ into $(s'-s)/\eps$ points separated by distance $\eps$. One then obtains 
\beq
I_\eps(s,s')=\int dv_i \prod_i\de(v_i^2-1)(\eps\sum v_i)^{b}.
\label{Isseps}
\eeq
This integral will, of course, vanish for symmetry reasons unless $b$ is even.
Furthermore, when we expand $(\eps\sum v_i)^{b}$, any terms containing odd powers of $v_i$ will vanish after the integration. Hence, the structure of the expanded $b$-power of the sum will be as follows: there will be $((s'-s)/\eps)^{b/2}$ terms which are products of $b/2$ values of $v_i^2$ at different points. For any terms with a higher power of $v_i$,
one will obtain a smaller number of terms ($\,(s'-s)/\eps$ to a power less than $b/2$) as
more points have to coincide while picking $v_i$'s from the sum. Hence such terms will
come with a greater power of $\eps$ (than the terms containing only $v_i^2$) and will not contribute to (\ref{Isseps}) in the $\eps\to 0$ limit. Therefore, the leading small $\eps$ behavior of (\ref{Isseps}) is
\beq
I_\eps(s,s')=\left(\eps(s'-s)\langle v^2\rangle\right)^{b/2},
\label{Iv2}
\eeq
where $\langle v^2\rangle\equiv\int dv v^2 \de(v^2-1)=1$. (As already mentioned under (\ref{Brwilson}), to obtain a continuum limit one needs to scale positions on the random walk as $s\sim 1/\eps$, which includes scaling the total length of the trajectory as $L_0\sim 1/\eps$.)

Note that (\ref{Iv2}) is very insensitive to the shape of the distribution of $v$ (as is typical of the situations in the spirit of the central limit theorem): any other distribution (different from $\de(v^2-1)$) with the same $\langle v^2\rangle$ will give the same result for (\ref{Iv2}), and hence for expectation values of any quantities of the type (\ref{F}). In particular, we can replace $\de(v^2-1)$ by a Gaussian distribution $\exp(-v^2)$ (again, the normalization of the measure does not concern us here). In the continuum limit, $\prod\exp(-v_i^2)$ becomes $\exp\left(-\int ds v^2/\eps\right)$, 
which is in agreement with the continuum Gaussian measure obtained in \cite{polyakov}.
The conclusion is that, for any quantity of the type (\ref{F}), the random walk measure
can be replaced as
\beq
\int d\xi(s)\,\de[\dot\xi^2-1]\to\int  d\xi(s) \exp\left[-\int ds\,\dot\xi^2/\eps\right],
\label{BrGauss}
\eeq
without affecting the continuum limit (i.e., a random walk with a fixed speed becomes replaced by a walk with a Gaussian distribution of speeds, which is much more convenient to deal with in a path integral).

Even though the randomized Wilson loop (\ref{Brwilson}) is not, strictly speaking, of the form (\ref{F}) as the different integrations of $\dot\xi$ are entangled with each other,
a similar argument (though more cumbersome to typeset) leads to the conclusion
that the integration measure in (\ref{Brwilson}) can be replaced as in (\ref{BrGauss}) as well, without affecting the continuum limit. (The key idea is again that, after discretization, the only terms contributing to the continuum limit are products of $v_i^2$ at different points, and terms containing higher powers of $v_i$ are suppressed by powers of $\eps$. Hence, only the expectation value of $v^2$ contributes to the continuum limit, and the (factorized) measure can be replaced by any other (factorized) measure with the same expectation value of $v^2$. We use this freedom to replace the integration measure
by the Gaussian one.)

In this section, we are considering $\xi(s)$ to be entirely unconstrained.
In this case, the end point of the `loop' does not have to coincide with the starting one.
In a gauge theory, such a Wilson `loop' would have to vanish by gauge invariance.
However, in Eguchi-Kawai models this vanishing does not occur due to the breakdown of the $R^D$ symmetry. Hence, there is no straightforward relation between computations with closed and unconstrained Wilson loops (the situation may be better in more sophisticated reduced models, where open Wilson loops would have to vanish just as they do in gauge theory, and would thus give a zero contribution to the sum over all loops). Nevertheless, we shall pursue
the computation of the unconstrained average to develop the relevant techniques, as this
case enjoys a great algebraic simplicity. We shall follow up with a consideration
of a sum over closed loops in the next section.

\begin{figure}[t]
\begin{center}
\vspace{2cm}
\epsfig{file=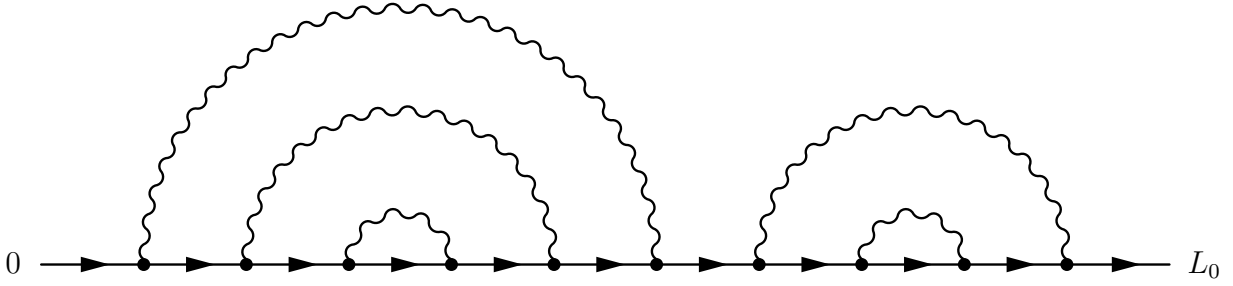}
\begin{picture}(0,0)
\put(0,25){$\,L_0$}
\put(-445,25){0}
\end{picture}
\end{center}
\vspace{-1.2cm}
\caption{Example of a rainbow diagram.}
\label{rainbow}
\end{figure}

Before we proceed evaluating the averaged Wilson loop (\ref{Brwilson}), we shall consider
the leading large $N$ large $D$ behavior of
\beq
a_{\mu_1\cdots\mu_{2n}}=\int dA_\mu^a e^{-S[A]} A_{\mu_1}^{a_1}\cdots A_{\mu_{2n}}^{a_{2n}} \frac1N\tr\left[t_{a_1}\cdots t_{a_{2n}}\right].
\label{Aintegral}
\eeq
Contracting all the $A$'s according to (\ref{Acontract}), we obtain
\beq
a_{\mu_1\cdots\mu_{2n}}=\sum\limits_{\{i(r)j(r)\}} \left(\frac{g}{\sqrt{2ND}}\right)^n\frac1N\tr\left[t_{a_1}\cdots t_{a_{2n}}\right]\prod\limits_{r=1}^n \de_{\mu_{i(r)}\mu_{j(r)}}\de_{a_{i(r)}a_{j(r)}}.
\label{largeDa}
\eeq
Here, we have introduced pairings $\{i(r),j(r)\}$ that for each pair number $r$ give the
two members of the pair $i(r)$ and $j(r)$, and the sum is over all such pairings.
It is well-known from the usual considerations of the large $N$ limit of gauge theories (see, e.g., \cite{makeenko,polyakov})
that the leading large $N$ contribution into $\tr\left[t_{a_1}\cdots t_{a_{2n}}\right]/N$
comes from planar contractions and it is equal to $N^n$ for any such planar contraction.
`Planar' in our context means that if all $t_a$'s are represented as dots on a horizontal straight line, and all the contractions $\de_{a_{i(r)}a_{j(r)}}$ in (\ref{largeDa}) as
arcs in the upper half plane connecting the corresponding points, such a diagram can
be drawn without any intersections. The result can be represented pictorially as `rainbow diagrams' or `arches' of the type depicted in Fig.~\ref{rainbow}. As only such diagrams are retained in the large $N$ limit,
one obtains
\beq
a_{\mu_1\cdots\mu_{2n}}=\sum\limits_{\mbox{\small planar}\{i(r)j(r)\}} \left(\frac{g\sqrt{N}}{\sqrt{2D}}\right)^n\prod\limits_{r=1}^n \de_{\mu_{i(r)}\mu_{j(r)}}.
\label{largeNa}
\eeq

We now proceed evaluating the randomized Wilson loop. From (\ref{Brwilson}), (\ref{BrGauss})
and (\ref{largeNa}),
\beq
\begin{array}{l}
\dsty W_{Br}(L)=
\int  d\xi(s)\,e^{-\int ds\,\dot\xi^2/\eps}\sum\limits_{n=0}^\infty (-1)^n \int\limits_0^{L^2/\eps} ds_1\cdots \int\limits_{s_{2n-1}}^{L^2/\eps}ds_{2n}\,\, a_{\mu_1\cdots\mu_{2n}}\prod\limits_{k=1}^{2n} \dot\xi^{\mu_k}(s_k)\vspace{3mm}\\
\dsty\hspace{1cm}=\int  d\xi(s)\,e^{-\int ds\,\dot\xi^2/\eps}\sum\limits_{n=0}^\infty \left(-\frac{g\sqrt{N}}{\sqrt{2D}}\right)^n \int\limits_0^{L^2/\eps} ds_1\cdots \int\limits_{s_{2n-1}}^{L^2/\eps}ds_{2n}\vspace{2mm}\\ \dsty\hspace{3cm}\times\sum\limits_{\mbox{\small planar}\{i(r)j(r)\}} \prod\limits_{r=1}^n \de_{\mu_{i(r)}\mu_{j(r)}}\prod\limits_{k=1}^{2n} \dot\xi^{\mu_k}(s_k).
\end{array}
\label{WBrproc}
\eeq
The Gaussian integration over $\xi(s)$ now merely amounts to contracting $\dot\xi_\mu$'s with
\beq
\langle\dot\xi^{\mu_1}(s_1)\dot\xi^{\mu_2}(s_2)\rangle = \eps \de^{\mu_1\mu_2} \de(s_1-s_2).
\label{xicontract}
\eeq
In order to maximize the power of $D$ in each term, $\dot\xi_\mu$'s should be contracted using exactly the same pairings $\{i(r),j(r)\}$ that have been used for contracting
$A$'s in (\ref{Aintegral}). With such a match, each $\de^{\mu_k\mu_l}$ from the contraction
(\ref{xicontract}) will hit the corresponding $\de_{\mu_{i(r)}\mu_{j(r)}}$ in (\ref{WBrproc})
producing a factor of $D$. For any other contraction pattern, the number of powers of $D$
will always be smaller, rendering such terms irrelevant in the large $D$ limit.
Bearing these observations in mind, we obtain
\beq
\dsty W_{Br}(L)=
\sum\limits_{n=0}^\infty \left(-\eps\sqrt\frac\lambda2\right)^n \int\limits_0^{L^2/\eps} ds_1\cdots \int\limits_{s_{2n-1}}^{L^2/\eps}ds_{2n} \sum\limits_{\mbox{\small planar}\{i(r)j(r)\}}\prod\limits_{r=1}^n \de(s_{i(r)}-s_{j(r)}),
\label{WBrde}
\eeq
where we have introduced the modified 't Hooft coupling $\lambda\equiv g^2ND$.

The rest of the computation becomes very simple now. The integration variables are arranged as
$s_1\le s_2\le\cdots\le s_{2n}$. A $\de$-function $\de(s_{i(r)}-s_{j(r)})$ enforces
$s_{i(r)}=s_{j(r)}$ and hence, all $s_i$'s with $i$ between $i(r)$ and $j(r)$ should be
equal to each other. Each such equality would make the integral vanish, except that
the singularity in the $\de$-function can compensate for equality of two of the $s_i$'s (but not more). That means that the integral in (\ref{WBrde}) will vanish unless all the pairings
are adjacent ($j(r)=i(r)+1$). (This property can be derived more cleanly by considering
a smeared $\de$-functions $\de(s)=\left\{1/2\Delta\mbox{ if }|s|<\Delta,\,\,\, 0\mbox{ otherwise}\right\}$ with $\Delta$ that has to be taken to $0$. One will then see that any integrals with non-adjacent pairings will be proportional to positive powers
of $\Delta$ and thus vanish in the limit recovering the $\de$-function proper.)

There is evidently only one rainbow diagram with only adjacent pairings for each $n$.
Furthermore,
\beq
\int\limits_0^{L^2/\eps} ds_1\cdots \int\limits_{s_{2n-1}}^{L^2/\eps}ds_{2n} \prod\limits_{r=1}^n \de(s_{2r-1}-s_{2r})=\frac1{2^nn!}\left(\frac{L^2}{\eps}\right)^n.
\eeq
Hence,
\beq
\dsty W_{Br}(L)=
\sum\limits_{n=0}^\infty \frac1{2^nn!}\left(-L^2\sqrt\frac\lambda2\right)^n=\exp\left(-\frac{L^2}{2}\sqrt\frac\lambda2\right).
\label{area}
\eeq
So at the end of the day, all the dependences on the random path integration cut-off $\eps$
have cancelled out (this simply reflects the fact that we have chosen the correct renormalization of the path length $L_0=L^2/\eps$ to maintain a finite spatial extent
of the random walk), and we obtain a very attractive expression for the averaged Wilson
loop (since $L$ is the typical spatial extent of the random walk, $L^2$ is the typical
area of its cross-section).

As explained above, the derivations of this section are a toy example and do not bear
an immediate relationship to the gauge theory (because the Wilson loops are not closed,
and because of the spontaneous symmetry breakdown issues in the Eguchi-Kawai model).
However, one may hope that the attractive mathematical structure we have displayed
will survive in more realistic examples. Furthermore, in the next section, we shall
repeat the present computation for the more realistic case of closed Wilson loops
and derive (with considerably more effort) a somewhat more complicated formula
resembling (\ref{area}).

\section{Closed Wilson loops}

We shall now constrain the Wilson loop trajectory $\xi(s)$ to start and end at 0.
This results in a physically meaningful summation over gauge-invariant closed Wilson loops.

Up to (\ref{WBrproc}), the derivation is exactly the same as in the previous section.\footnote{Technically, for closed loops, $\dot\xi(s)$ are not completely independent at different points, as there is one global constraint $\int_0^{L_0}ds\,\dot\xi=0$. This constraint, however, does not upset the application of the central limit theorem. Indeed, one can make different $\dot\xi(s)$ completely uncorrelated by explicitly inserting $\de(\int_0^{L_0}ds\,\dot\xi)=\int dk \exp\left(ik\int_0^{L_0}ds\,\dot\xi\right)$. With an unconstrained integration over $\dot\xi(s)$, one can consider again the expectation value of the type (\ref{F}) and include $\int dk \exp\left(ik\int_0^{L_0}ds\,\dot\xi\right)$ as a part of $F$. The derivations between (\ref{F}) and (\ref{BrGauss}) will then apply, the measure will become replaced by the Gaussian one, after which one restores $\de(\int_0^{L_0}ds\,\dot\xi)$ and removes it re-introducing the Dirichlet boundary conditions $\xi(0)=\xi(L_0)=0$. The result is the original closed loop integral, but with the $\de$-function measure replaced by the Gaussian measure, as in (\ref{BrGauss}). Again, a variation of this argument will apply to the expectation value of the Wilson loop, which is not, strictly speaking, of the form (\ref{F}).}
Thereafter, to contract the $\xi$'s, one needs to use the Dirichlet Green function
on a line:
\bea
&\dsty\langle\xi^{\mu_1}(s_1)\,\xi^{\mu_2}(s_2)\rangle = \eps \de^{\mu_1\mu_2} D(s_1,s_2),\\
&\dsty D(s_1,s_2)=\mbox{min}(s_1,s_2)\left(1-\frac{\mbox{max}(s_1,s_2)}{L_0}\right).\rule{0mm}{7mm}
\eea
(Once again, $L_0=L^2/\eps$.) Or, for $\dot\xi$:
\bea
&\dsty\langle\dot\xi^{\mu_1}(s_1)\dot\xi^{\mu_2}(s_2)\rangle = \eps \de^{\mu_1\mu_2} {\cal D}(s_1,s_2),\\
&\dsty{\cal D}(s_1,s_2)\equiv\del_{s_1}\del_{s_2}D(s_1,s_2)=\de(s_1-s_2)-\frac{1}{L_0}.\rule{0mm}{7mm}
\label{Dirder}
\eea
Then,
\beq
\dsty W_{Br}(L)=
\sum\limits_{n=0}^\infty \left(-\eps\sqrt\frac\lambda2\right)^n \int\limits_0^{L^2/\eps} ds_1\cdots \int\limits_{s_{2n-1}}^{L^2/\eps}ds_{2n} \sum\limits_{\mbox{\small planar}\{i(r)j(r)\}}\prod\limits_{r=1}^n {\cal D}(s_{i(r)},s_{j(r)}).
\eeq
This expression only differs from (\ref{WBrde}) by the second (constant) term on the right-hand side of (\ref{Dirder}). Yet, such a minor difference turns out to complicate
the remaining derivation considerably. We shall first re-scale $s$ so that it runs from 0 to 1:
\beq
\dsty W_{Br}(L)=
\sum\limits_{n=0}^\infty \left(-L^2\sqrt\frac\lambda2\right)^n \int\limits_0^{1} ds_1\cdots \int\limits_{s_{2n-1}}^{1}ds_{2n} \sum\limits_{\mbox{\small planar}\{i(r)j(r)\}}\prod\limits_{r=1}^n \left(\de(s_{i(r)}-s_{j(r)})-1\right).
\label{WBr1}
\eeq
One sees that the random walk integration cut-off $\eps$ disappears as it should.
Expanding the product in the above formula, one will have a collection of terms with
different numbers of $\de$-functions. As explained under (\ref{WBrde}), these integrals
will vanish, unless all the pairings appearing in the $\de$-function are adjacent ($j(r)=i(r)+1$). If that is so, a general term of this type will have $k$ $\de$-functions
whose argument is a difference of adjacent $s_i$'s (denoted $s_{p(l)}$ and $s_{p(l)+1}$), and it can be evaluated as
\beq
\int\limits_0^{1} ds_1\cdots \int\limits_{s_{2n-1}}^{1}ds_{2n} \prod\limits_{l=1}^k \de(s_{p(l)}-s_{p(l)+1}) =\frac1{2^{k}(2n-k)!}.
\label{deint}
\eeq

We now introduce $R_{mn}$, the number of rainbow diagrams (Fig.~\ref{rainbow}) with $m$ adjacent and $n$ non-adjacent pairings, and consider a rainbow diagram with a total of $n$ points, $m$ of which are adjacently-paired. For all non-adjacently-paired points, the $\de$-function
term in (\ref{WBr1}) cannot contribute. For the adjacently-paired points, one can have contributions from both $\de$-function and constant term. Furthermore, for $k$ $\de$-functions, there are $C_m^k$ ways to choose their positions on the $m$ arches connecting adjacent points,
all resulting in the same value of (\ref{deint}). Hence, (\ref{WBr1}) can be re-written using (\ref{deint}) as
\beq
\begin{array}{l}
\dsty W_{Br}(L)=
\sum\limits_{n=0}^\infty \left(-L^2\sqrt\frac\lambda2\right)^n \sum\limits_{m=0}^n R_{m,n-m}\sum\limits_{k=0}^m \frac{(-1)^{n-k}C^k_m}{2^{k}(2n-k)!}\vspace{3mm}\\
\dsty\hspace{1cm}=\sum\limits_{m,n=0}^\infty R_{mn}\left(L^2\sqrt\frac\lambda2\right)^{n+m}\sum\limits_{k=0}^m \frac{(-1)^{k}C^k_m}{2^{k}(2(n+m)-k)!}.
\end{array}
\label{WBrrainbow}
\eeq
What remains now is to evaluate the rainbow numbers $R_{mn}$ and to resum (\ref{WBrrainbow}).

The rainbow numbers satisfy a recursion relation
\beq
R_{m+1,n+1}=R_{m,n+1}+\sum\limits_{k=1}^{m+1}\sum\limits_{l=0}^{n}R_{kl}R_{m+1-k,n-l}
\label{recur}
\eeq
with the boundary conditions
\beq
R_{n0}=1,\quad R_{1n}=1\,\,\,(\mbox{any } n),\quad R_{0n}=0\,\,\,(n>0).
\label{bdry}
\eeq
This recursion relation can be easily understood: consider a rainbow with $m+1$ adjacent
and $n+1$ non-adjacent pairings. If the first point is connected to the second one (i.e., adjacently paired),
the remaining diagram is a rainbow with $m$ adjacent and $n+1$ non-adjacent pairings.
This gives the first term. Alternatively, the first point can be connected to another
point with an even number. Then, the arch coming out of the first point will split
the diagram into two smaller rainbow diagrams (under this arch and outside this arch)
with a total number of $m+1$ adjacent and $n$ non-adjacent pairings. This gives the second term.

The recursion relation (\ref{recur}) is closely reminiscent of the one for Catalan numbers.
This is not surprising, since the total number of rainbows $R_n=\sum_m R_{m,n-m}$ is precisely
given by Catalan numbers (see p. 222-223 of \cite{makeenko}). To solve the recursion relation,
one introduces a generating function
\beq
R(x,y)=\sum\limits_{m,n=0}^\infty R_{mn}x^my^n,
\label{gener}
\eeq
which allows to rewrite (\ref{recur}) and (\ref{bdry}) as
\beq
\frac1{xy}\left(R-\frac{1}{1-x}\right)=\frac1{y}\left(R-\frac{1}{1-x}\right)+\frac{R(R-1)}{x}.
\eeq
This quadratic equation is solved by
\beq
R=-\frac{x-y-1}{2y}\pm\sqrt{\left(\frac{x-y-1}{2y}\right)^2-\frac1y}.
\eeq
Keeping (\ref{gener}) in mind, we must choose a solution non-singular at the origin,
which is the one with a minus sign. This non-singular solution can also be re-written as
\beq
R=\frac{2}{1-x+y+\sqrt{(1-x+y)^2-4y}}.
\eeq
As a cross-check, the generating function for the Catalan numbers $(1-\sqrt{1-4x})/2$ should
result from setting $x=y$, which is indeed the case. The first few terms of the power series for $R$ are
\beq
\begin{array}{l}
\dsty R=1+x+x^2+xy+x^3+3x^2y+xy^2+x^4+6x^3y+6x^2y^2+xy^3\vspace{2mm}\\
\dsty\hspace{4cm}+x^5+10x^4y+20x^3y^2+10x^2y^3+xy^4+\cdots,
\end{array}
\eeq
and they can be verified to match the numbers of low order rainbow diagrams.

A brute force way to proceed with the computation would now be to extract the rainbow
numbers by expanding $R$ and then try to resum (\ref{WBr1}). However, one can do wiser
than that and attempt to convert the (already resummed) expression for $R$ into an expression
for the Wilson loop. To this end, we observe that
\beq
R\left(x\left(1-\frac{z}{2}\right),y\right)=\sum\limits_{m,n=0}^\infty R_{mn}x^my^n\sum\limits_{k=0}^m C^k_m \left(-\frac{z}{2}\right)^k
\eeq
and
\beq
R\left(\frac{x}{z^2}\left(1-\frac{z}{2}\right),\frac{y}{z^2}\right)=\sum\limits_{m,n=0}^\infty R_{mn}x^my^n\sum\limits_{k=0}^m C^k_m \left(-\frac{1}{2}\right)^k z^{-2(m+n)+k}.
\eeq
Furthermore, using
\beq
\frac1{2\pi i}\int\limits_{c-i\infty}^{c+i\infty} dz \frac{e^z}{z^{n+1}}=\frac1{n!}
\eeq
(where $c$ is a positive number), we obtain
\beq
\frac1{2\pi i}\int\limits_{c-i\infty}^{c+i\infty} dz\,\frac{e^z}{z} R\left(\frac{x}{z^2}\left(1-\frac{z}{2}\right),\frac{y}{z^2}\right)=\sum\limits_{m,n=0}^\infty R_{mn}x^my^n\sum\limits_{k=0}^m \frac{(-1)^{k}C^k_m}{2^{k}(2(n+m)-k)!}.
\eeq
Here, we immediately recognize a structure matching (\ref{WBr1}). Hence, with $\ell^2\equiv L^2\sqrt{\lambda/2}$,
\beq
\begin{array}{l}
\dsty 
W_{Br}(L)=\frac1{2\pi i}\int\limits_{c-i\infty}^{c+i\infty} dz\,\frac{e^z}{z} R\left(\frac{\ell^2}{z^2}\left(1-\frac{z}{2}\right),\frac{\ell^2}{z^2}\right)
\vspace{3mm}\\
\dsty\hspace{3cm}
=\frac1{2\pi i}\int\limits_{c-i\infty}^{c+i\infty} dz\,\frac{4e^z}{\dsty 2z+\ell^2+\sqrt{\left(2z+\ell^2\right)^2-16\ell^2}}.
\end{array}
\eeq
Because of the fast decay of the exponential, the contour can actually be closed at $z=-\infty$ and then pulled over the cut of the square root in the integrand. We obtain
\beq
\begin{array}{l}
\dsty W_{Br}(L)=\frac{\exp\left(-\ell^2/2\right)}{\pi i}\int\limits_{c-i\infty}^{c+i\infty} dz\,\frac{e^z}{\dsty z+\sqrt{z^2-4\ell^2}}=\frac{\exp\left(-\ell^2/2\right)}{2\pi \ell^2}\int\limits_{-2\ell}^{2\ell}dx\,e^x\sqrt{4\ell^2-x^2}\vspace{3mm}\\
\dsty \hspace{4cm}=\exp\left(-\frac{L^2}{2}\sqrt\frac\lambda2\right)\,\,\frac{2}{\pi }\int\limits_0^\pi d\alpha\, e^{2L\left(\lambda/2\right)^{1/4}\cos\alpha}\sin^2\alpha.
\end{array}
\eeq
The leading behavior here is again in the form of an area law:
\beq
\lim\limits_{L\to\infty} \frac{\ln W_{Br}(L)}{L^2}=-\sqrt\frac{\lambda}{8}.
\eeq

\section{Conclusions}

Considering randomized Wilson loops (Wilson loops averaged over an ensemble of random
contours) in the context of Eguchi-Kawai reduced model results in an algebraically tractable dynamically non-trivial large $D$ limit ($D$ being the number of matrices
in the Eguchi-Kawai model, corresponding to the number of dimensions in gauge theory).
The scale dependence of the expectation values of such randomized Wilson loops
turns out to be strongly reminiscent of the area law.

It would be interesting to generalize the present considerations for randomized Wilson
loops to more realistic reduced models \cite{Azeyanagi:2010ne,current}. This should
be possible once the large $D$ limit has been worked out for those theories.
The area law seems to emerge quite naturally from the algebraic structures
appearing in the randomized Wilson loop computations (a series in powers of $L^2$, i. e.,
the area, with alternating signs and factorial denominators). It will not be surprising
if analogous computations for more realistic cases display an area law as well.

It is also interesting to speculate how much the behavior of randomized Wilson loops
constrains the behavior of individual Wilson loops. The area law decrease for large Wilson
loops is very fast, and having individual Wilson loops decreasing slower than the area law
would seem in conflict with the fast decrease of the randomized average. Should it be
possible to show that expectation values of individual Wilson loops are positive,
the fast decrease of the randomized average will become even more constraining
(if a sum of positive numbers is small, it constrains the magnitude of individual
entries in the sum). A better understanding of the relations between randomized and
individual Wilson loop averages could also help to establish the randomized Wilson
loop analyzed here as an order parameter for the confining phase of the gauge theory
(cf. the preceding remarks on the area law behavior). These issues certainly deserve further investigation.

Since the matrix integral (\ref{originalaction}) appears as a small volume limit
of compactified gauge theory (irrespectively of the Eguchi-Kawai equivalence), our
present derivations formally apply to the study of that limit as well. 
The caveat is that the
matrix model Wilson loop (\ref{wilson}) does not appear to have special significance,
unless it can be related to the gauge theory Wilson loop via the Eguchi-Kawai equivalence.
Some applications of the high temperature/small volume limit of gauge theories
can be found in \cite{ht}.

It is instructive to examine our results in the light of numerical findings reported in \cite{arealaw}. In that publication, area law for Wilson loop expectation values has been
seen via numerical simulations for sufficiently large rectangular Wilson loops in bosonic and supersymmetric matrix models similar to (\ref{originalaction}). This observation was
perceived as an evidence that the Eguchi-Kawai equivalence holds for sufficiently large spatial scales
(since the corresponding gauge theories are confining and should indeed exhibit the area law). The validity of the Eguchi-Kawai equivalence for sufficiently large scales has
been further heuristically linked to another numerical result described in \cite{arealaw}: the smallness of Polyakov lines of sufficiently large circumference. Since vanishing of Polyakov lines would imply a lack of spontaneous breaking of the $U(N)$ center symmetry and hence exact validity of the Eguchi-Kawai equivalence, the observed smallness of Polyakov lines was seen as an additional indication of approximate validity of the Eguchi-Kawai equivalence for sufficiently large spatial scales (physically, above the characteristic size of confined one-particle states). All these observations mesh quite well with the derivations presented here: if the heuristics of \cite{arealaw} is taken one step further, one may hypothesize that the Eguchi-Kawai equivalence applies to
randomized Wilson loops of sufficiently large characteristic size ($L$). Then the present results will have direct implications for gauge theory irrespectively of spontaneous
breaking of the $U(N)$ center symmetry that undermines exactness of the Eguchi-Kawai equivalence. Of course, such claims are far from high standards of rigor, as are those of \cite{arealaw}. Whatever the case, the derivations presented here can be seen as a first step towards analytic understanding of the numerical observations reported in \cite{arealaw}.

\section{Acknowledgments}
 
I would like to thank Gautam Mandal, Takeshi Morita, Takuya Okuda and Mithat \"Unsal for valuable discussions. This research has been
supported by grants from the Chinese Academy of Sciences and National Natural Science Foundation of China.


\end{document}